\documentclass[conference]{IEEEtran}

\input epsf
\usepackage{graphicx}
\usepackage{multirow}
\usepackage{hhline}
\usepackage{diagbox}
\usepackage{amsmath}
\usepackage[caption=false]{subfig}
\usepackage{algpseudocode,algorithm,algorithmicx}

\begin{document}

\title{Exploring Erasure Coding Techniques for High Availability of Intermediate Data}
\author{\authorblockN{Zhe Zhang\IEEEauthorrefmark{1}, Brian Bockelman\IEEEauthorrefmark{2}, Derek Weitzel\IEEEauthorrefmark{1}, David Swanson\IEEEauthorrefmark{1}\\ zhan0915@huskers.unl.edu, bbockelman@morgridge.org, dweitzel@cse.unl.edu, dswanson@cse.unl.edu}
\authorblockA{\IEEEauthorrefmark{1}Holland Computing Center, University of Nebraska $-$ Lincoln, Lincoln, NE 68588, USA}
\authorblockA{\IEEEauthorrefmark{2}Morgridge Institute for Research, Madison, WI 53715, USA}
}




%


\maketitle

\begin{abstract}
Scientific computing workflows generate enormous distributed data that is short-lived, yet critical for job completion time. This class of data is called intermediate data. A common way to achieve high data availability is to replicate data. However, an increasing scale of intermediate data generated in modern scientific applications demands new storage techniques to improve storage efficiency. Erasure Codes, as an alternative, can use less storage space while maintaining similar data availability. In this paper, we adopt erasure codes for storing intermediate data and compare its performance with replication. We also use the metric of Mean-Time-To-Data-Loss (MTTDL) to estimate the lifetime of intermediate data. We propose an algorithm to proactively relocate data redundancy from vulnerable machines to reliable ones to improve data availability with some extra network overhead. Furthermore, we propose an algorithm to assign redundancy units of data physically close to each other on the network to reduce the network bandwidth for reconstructing data when it is being accessed.

\end{abstract}
\IEEEoverridecommandlockouts

%
\IEEEpeerreviewmaketitle
\section{Introduction}
\label{sec:introduction}

Scientific workflow is one of the most popular ways to map scientific applications to computational resources. A scientific workflow is usually represented by a Direct Acyclic Graph (DAG). In such a graph, a task represented as a vertex is connected with other tasks. An output data generated from a task needs to be read as an input to one or multiple succeeding tasks. This class of data is called \textit{intermediate data}.

As today's scientific applications become more and more complex, enormous intermediate data can be generated from a scientific workflow \cite{evolution}. For example, LIGO is a network of gravitational-wave detectors. In the LIGO project, a meaningful run of binary inspiral workflows requires a minimum of 221 GB of gravitational-wave data and approximately 70,000 computational tasks \cite{optimizefootprint}. The aggregate intermediate data can easily go to the Petabyte scale. Managing these data becomes a challenging problem for the system implementers \cite{ligodata}.

One of the challenges is the frequent failures in modern scientific grids. For example, in the Open Science Grid (OSG) - one of the state-of-the-art scientific grids, 40\% of pilots (similar to virtual machines in cloud computing) \cite{pilotsystem} encounter failures \cite{Zhang:2018:DJP:3219104.3229282}. Data loss of intermediate data can halt the workflow and result in a significant delay in job completion.

Two approaches are commonly used to recover intermediate data. The first approach is to replicate data to multiple storage resources to overcome single resource failure \cite{gfs}\cite{hdfs}. Ko \cite{intermediate} suggested replicating locally consumed intermediate data to remote machines. In the case of a node failure, a succeeding task can be restarted on a node where the input data is available. Jeon \cite{cloudbackup} proposed to leverage Cloud storage to back up intermediate data. However, those methods did not account for the storage limit that modern grid systems face \cite{limit}. In addition, storing intermediate data in Cloud introduces long latency to accessing data, which is unacceptable to some time-critical applications \cite{timecritical}. Some scientific organizations offer dedicated storage servers to store user data. StashCache \cite{stashcache} which is implemented in the OSG offers a few geographically distributed storage servers. It allows a user to select a server to cache intermediate data. However, it relies on high-end facilities in the infrastructure. It also exposes complexity to a user to deploy the application to the system.

Another approach to recover intermediate data is to recompute tasks \cite{Deelman201517}\cite{mapreduce}. This approach sometimes can result in what we called cascaded re-execution: some tasks in every stage from the beginning have to be re-executed sequentially up to the stage where the failure happened. This is one of the most common reasons for job delays in data centers \cite{intermediate}. Although some research tries to optimize the recomputing cost by persisting or reusing parent intermediate data \cite{rdd}\cite{recomputation}, the frequent failure appearance in opportunistic scientific grid systems \cite{opportunistic}\cite{Zhang:2018:DJP:3219104.3229282} makes it not an appropriate solution. In a short, data replication pays storage cost for data availability; task recomputation pays computing power instead. Mantri \cite{mantri} proposed a model to evaluate the cost to recompute intermediate data. If the cost goes beyond a certain threshold, it automatically replicates data.

In this paper, we explore erasure coding techniques for intermediate data. As an alternative to data replication, erasure codes can save storage space. Compared with task recomputation, it does not generate extra computation cost. Although erasure codes have been well studied, the existing research mainly focuses on permanent storage - the majority of the data is rarely accessed after being stored. Applying erasure codes to intermediate data is barely addressed. The main reason is due to high network bandwidth for reconstructing data \cite{regeneratingcode} when data is being accessed. However, in modern scientific grid systems in which virtualization techniques have been broadly adopted, this effect can be mitigated. We can assign data chunks physically close to each other in hardware but logically isolated in a virtual layer to reduce the network traffic. This paper has three contributions to the community of scientific computing:

\begin{itemize}
\item
We implemented erasure codes in HTCondor \cite{condorpractice} - a widely adopted software tool in scientific computing. We compare erasure codes with data replication in the following system metrics: storage cost, data availability and network bandwidth cost.
\item
We propose a proactive algorithm to improve data availability for intermediate data. By using the Mean-Time-To-Data-Loss (MTTDL) model \cite{raid}, the algorithm estimates the lifetime of intermediate data. If data is approaching the end of its life, the algorithm automatically relocates data chunks to prolong the data's life.
\item
We propose an algorithm to assign data chunks of intermediate data physically close to each other on the network to reduce network traffic for reconstructing data.
\end{itemize}
\section{Background}
\label{sec:background}

\subsection{Intermediate Data}
\label{subsec:intermediate}

We gleaned two main characteristics of intermediate data from scientific applications. First, unlike persistent data that typically leverage distributed file system \cite{hdfs} or object storage \cite{amazon}, most of the intermediate data relies on local storage \cite{mapreduce}\cite{Deelman201517}. Running out of disk quota is one of the main reasons that cause jobs to fail \cite{taobao}. Thus, minimizing storage space is an important task to improve job reliability \cite{optimizefootprint}. Second, persistent data is long-lived. In contrast, intermediate data in scientific workflows is short-lived because once a block is written by a task, it is transferred to and used immediately by the next task. Estimating the lifespan of intermediate data can be useful for improving system utilization: on one hand, if data is lost before the next tasks read it, recomputation will occur; on the other hand, if data exists in the system for an unnecessarily long period, it can cause a storage waste. We will discuss how to estimate intermediate data lifetime in Section \ref{subsec:markov}.

\subsection{Erasure Codes}
\label{subsec:ec}

Erasure coding can be viewed as an operation that takes $k$ units of data and generates $n = (k + r)$ units of data that are functions of the original $k$ data units. Typically, in the codes employed in storage systems, the first $k$ of the resultant $n$ units are identical to the original $k$ units. These units are called data units. The $r$ additional units generated are called parity units. The parity units are some mathematical functions of the data units, and thus contain redundant information associated with the data units. This set of $n = (k + r)$ units is called a stripe. In this paper, we call both data units and parity units as \textit{redundancy units}.

The redundancy of a code is defined as the ratio of the stripe size to the logical size of the original data:

\begin{equation}
Redundancy = \frac{n}{k}
\end{equation}

In this paper, we use \textit{Replica(n)} to represent a replication policy with n copies; \textit{EC(k+r)} to represent a erasure coding policy with k data units and r parity units. For example, Replica2 has a redundancy of 2 and EC2+1 has a redundancy of 1.5.

\subsection{System Reliability}
\label{subsec:fr}

In system reliability engineering, Weibull distribution is commonly used to model system failures \cite{weibull1}. We adopt the model presented in \cite{Litke:2007:ETR:1232286.1232287} to quantify failure probability:

we assume that a task is going to start its execution at time $t_0$ that assumes a pilot which carries the task is alive at time $t_0$. The failure rate of the task, as expressed in Equation \ref{eq:eq5}, is defined as the probability of failure during the next $\Delta t$ time units. 

\begin{equation}
\label{eq:eq5}
P(t_0<s<t_0+\Delta t|s>t_0) = P(t_0<s<t_0+\Delta t)/P(s>t_0)
\end{equation}

where $s$ represents the time that the task fails.

We use $p(t)$ to represent system lifetime probability density function and $f$ to represent the failure rate of the task, then the above equation can be further represented as follows:

\begin{equation}
\label{eq:failurerate}
f = \int_{t_0}^{t_0+\Delta t}p(t) / \int_{t_0}^{\infty}p(t)
\end{equation}

\subsection{Markov Model for MTTDL}
\label{subsec:markov}

MTTDL has been the standard reliability metric in storage systems for more than 20 years \cite{raid}. MTTDL represents a simple formula that can be used to estimate a time period that data can be retained in a storage system. The data loss rate is the inverse of MTTDL. We use both metrics to evaluate data availability in this paper.

\begin{figure}[ht]
\begin{center}
\includegraphics[height=0.5in, width=2.0in]{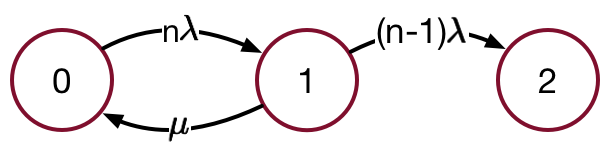}
\end{center}
\vspace*{-5mm}
\caption{Markov model on RAID5}
\label{fig:markovraid5}
\end{figure}

Figure \ref{fig:markovraid5} shows the Markov model for RAID5 which can tolerate single failure. There are a total of three states. State 0 is the state with all n redundancy units available. State 1 is the state with one lost redundancy unit. State 2 is the state with two lost redundancy units which also means data loss occurs in this state. The model in Figure \ref{fig:markovraid5} has two rate parameters: $\lambda$ - failure rate which indicates how frequently a failure appears in a system; $\mu$ - repair rate which represents how quickly a system can recover a failure. It is assumed that all devices fail at the same rate and repair at the same rate. By solving the absorbing Markov model, we can get the closed-form MTTDL in Equation \ref{eq:mttdlraid5}.

\begin{figure}[ht]
\begin{center}
\includegraphics[height=0.6in, width=2.6in]{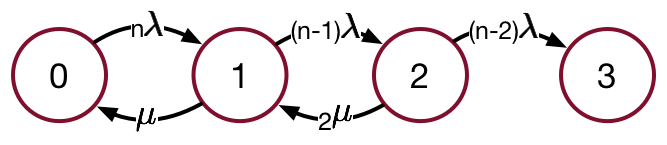}
\end{center}
\vspace*{-5mm}
\caption{Markov model on RAID6}
\label{fig:markovraid6}
\end{figure}

As shown in Figure \ref{fig:markovraid6}, RAID6 has four states. Therefore, it can tolerate two concurrent failures. The MTTDL of RAID6 is shown in Equation \ref{eq:mttdlraid6}.

\begin{equation}
\label{eq:t0raid5}
t_0 = \frac{1}{(n-1)\lambda}
\end{equation}

\begin{equation}
\label{eq:t1raid5}
t_1 = \frac{1}{n\lambda} + \frac{\mu}{n(n-1)\lambda^2}
\end{equation}

\begin{equation}
\label{eq:mttdlraid5}
MTTDL = t_0 + t_1
\end{equation}

\begin{equation}
\label{eq:t0raid6}
t_0 = \frac{1}{(n-2)\lambda}
\end{equation}

\begin{equation}
\label{eq:t1raid6}
t_1 = \frac{1}{(n-1)\lambda} + \frac{2\mu}{(n-1)(n-2)\lambda^2}
\end{equation}

\begin{equation}
\label{eq:t2raid6}
t_2 = \frac{1}{n\lambda} + \frac{\mu}{n(n-1)\lambda^2} + \frac{2\mu^2}{n(n-1)(n-2)\lambda^3}
\end{equation}

\begin{equation}
\label{eq:mttdlraid6}
MTTDL = t_0 + t_1 + t_2
\end{equation}

We extend the MTTDL model to RAID\textbf{r}. This term is different from RAID5 and RAID6. \textbf{r} represents the maximum recovering capability. Thus, RAID5 is rephrased to RAID1 because it is able to recover one failure. RAID6 is interpreted to RAID2. Equation \ref{eq:mttdlraidr}-\ref{eq:numerator} shows the MTTDL of RAIDr.

\begin{equation}
\label{eq:mttdlraidr}
MTTDL = \sum_{i=0}^{r} t_i = \sum_{i=0}^{r} \sum_{j=0}^{i} \frac{N_j}{D_j}
\end{equation}

\begin{equation}
\label{eq:denominator}
D_j = \prod_{k=0}^{j} (n-(r-i+k))\lambda
\end{equation}

\begin{equation}
\label{eq:numerator}
N_j = 
\begin{cases}
1.0, j=0\\
\prod_{k=1}^{j} (r-i+k)\mu, j>0\\
\end{cases}
\end{equation}

The MTTDL model depends on four parameters: n, r, $\lambda$ and $\mu$. $\lambda$ and $\mu$ are specific parameters related to system implementation. We will address these two parameters in Section \ref{subsec:parameters}. n and r are related to storage policies. For example, EC3+2 has n and r set to 5 and 2.
\section{Methodology}
\label{sec:methodology}

We implemented erasure codes by integrating Jerasure library \cite{Plank07jerasure:a} to HTCondor. We use Reed-Solomon code in Jerasure to encode and decode data. We also rely on Witzel's CacheD framework \cite{cached} which was designed to distribute shared caches to multiple worker nodes in high throughput computing systems. A \textit{CacheD} is a daemon/process running on a machine, which can send and receive redundancy units to other CacheDs. We present some terminology that will appear in the rest of the paper. An intermediate data stored in our system is called a \textit{cache}. Redundancy units of a cache are calculated by Jerasure and distributed to a few CacheDs which form a \textit{CacheCluster}. One of the CacheD in a CacheCluster is selected as the \textit{CacheManager}. The rest of the CacheDs are called \textit{CacheWorkers}. A CacheManager maintains meta-information about the CacheCluster. CacheWorkers need to send heartbeats to their CacheManager. If a CacheManager loses connection from a CacheWorker for a certain period, it takes the responsibility for recovering the cache and distributing the lost redundancy unit to a new CacheD.

\subsection{Testbed}
\label{subsec:testbed}

Our system consists of 5 Virtual Machines (VMs). 1 VM is \textit{master VM} which interacts with the client and accepts tasks. The rest of 4 VMs are \textit{slave VMs}, each of which can spawn multiple CacheD daemons to store redundancy units. These 5 VMs form a physical cluster. This cluster should be differentiated from the CacheCluster we described above. CacheCluster is a logical cluster that only pertains to a specific cache. A CacheD is used to simulate a machine entity in a distributed system. The reliability model of CacheDs follows a Weibull distribution.

A client task is simply to download a cache from a central repository. When the client wants to execute a task in the cluster, it negotiates with the master VM and decides which CacheD should process the task. After the master VM chooses a CacheD, the client directly schedules the task to it. This CacheD is the CacheManager for this cache. The cluster needs to manage the data internally based on different storage policies. If a replication policy is used, the CacheManager needs to contact CacheWorkers to distribute the replicas. If an erasure coding policy is used, the CacheManager needs to take one more step - calculating data and parity units - before distributing those redundancy units to CacheWorkers.

\subsection{System Lifetime Generation}
\label{sec:reliability}

\begin{figure}[ht!]
     \centering
     \subfloat[Weibull distribution]{\includegraphics[height=1.2in, width=1.7in]{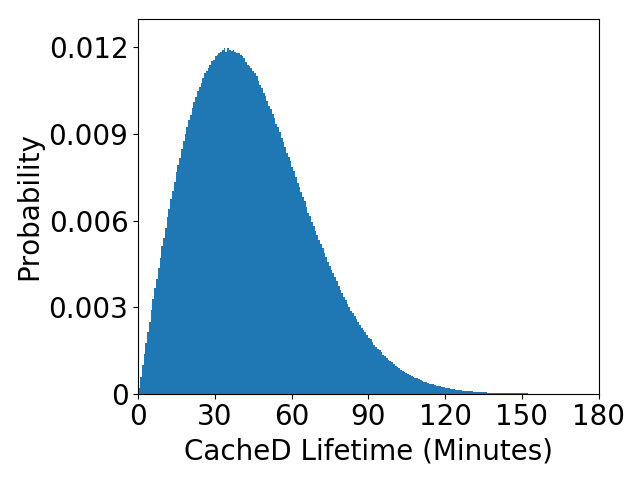}\label{fig:weibull}}
     \subfloat[Failure rate curve]{\includegraphics[height=1.2in, width=1.7in]{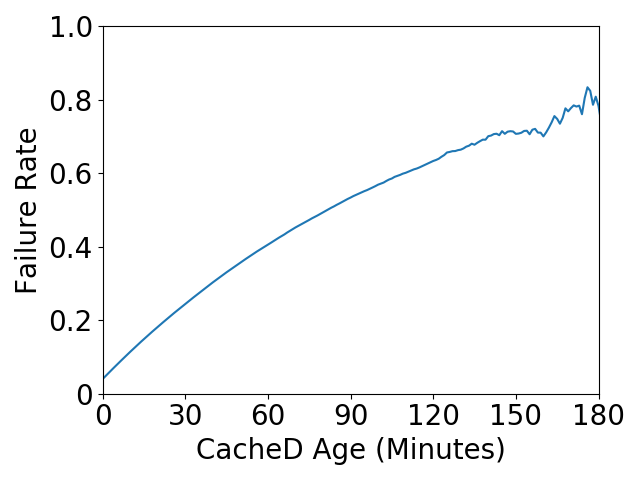}\label{fig:failurecurve}}
     \caption{Lifetime distribution and failure rate curve}
     \label{fig:pilotweibull}
\end{figure}

The master VM uses a Weibull distribution shown in Equation \ref{eq:weibull} to generate lifetime for CacheDs. In Equation \ref{eq:weibull}, $a$ is the shape parameter and $b$ is the scale parameter. In our experiments, we set $a = 2$ and $b = 50$. We generate 10 million samples by \textit{weibull\_min} function using \textit{scipy stats} library \cite{scipy}. The lifetime distribution is shown in Figure \ref{fig:weibull}. We define a parameter called \textit{lease period}. It is the expected period for a cache to stay in the system. We set it to 10 minutes in our experiments. Figure \ref{fig:failurecurve} shows failure rate curve calculated by Equation \ref{eq:failurerate} with $\Delta t = 10$ minutes.

\begin{equation}
\label{eq:weibull}
p(x) = \frac{a}{b} \cdot (\frac{x}{b})^{a-1} \cdot e^{-(\frac{x}{b})^a}, x \geq 0
\end{equation}

\subsection{Experiments}
\label{sec:experiments}

With the generator of the Weibull distribution, we used \textit{timeout} command in Linux to set a lifetime for a daemon when it got spawned. The client scheduled a task to the cluster every 30 seconds. We tested five storage policies: Replica1, Replica2, EC2+1, EC3+1, and EC3+2. In each test, we kept scheduling tasks for 120 minutes and therefore roughly 240 data were created in a test. If a cache had a sufficient number of redundancy units left for reconstructing the original data after the lease expired, we counted the cache as a success. Otherwise, we counted it as a data loss.

\subsection{Determining MTTDL parameters}
\label{subsec:parameters}

\begin{figure}[!ht]
\begin{center}
\includegraphics[height=1.8in, width=2.5in]{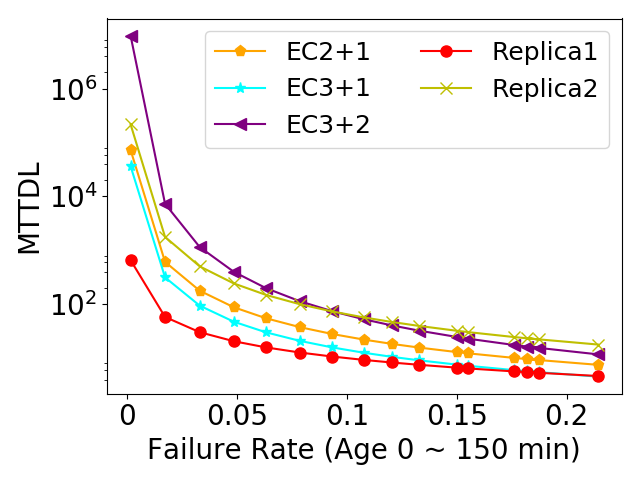}
\end{center}
\vspace*{-5mm}
\caption{MTTDL of different storage policies}
\label{fig:mttdlpolicies}
\end{figure}

As shown in Section \ref{subsec:markov}, MTTDL depends on four parameters: r, n, $\lambda$ and $\mu$. r and n are determined by storage policies. They will not change once a storage policy is chosen. The failure rate $\lambda$ and the repair rate $\mu$ vary depending on the system implementation. When a cache is deployed onto the cluster, the CacheManager periodically checks the availability of its CacheWorkers. If any CacheWorker has not sent a heartbeat for a certain period of time, the CacheManager marks the worker daemon as DOWN and starts to recover the failure. The time interval for the CacheManager to decide to recover the failure is a configurable variable in our system. In our experiment, we set it to 2 minutes. It means any failure during this interval cannot be recovered until the CacheManager checks the availability of the CacheWorkers. We use this interval as the finest granularity in our MTTDL model and set $\mu$ as 1. The failure rate of a CacheWorker should be estimated in the 2-minute interval. If we look at Section \ref{subsec:fr}, $p(t)$ follows our Weibull distribution and $\Delta t$ is 2 minutes. The only variable to calculate the failure rate is the age of a CacheWorker $t_0$. Figure \ref{fig:mttdlpolicies} shows how MTTDL changes with CacheD age from 0 to 150 minutes.

We can make a summary of three correlations between MTTDL and its four parameters from Equation \ref{eq:mttdlraidr} and Figure \ref{fig:mttdlpolicies}:

\begin{itemize}
\item
With other parameters fixed, MTTDL decreases as n increases. For example, EC3+1 has longer MTTDL than EC2+1.
\item
With other parameters fixed, MTTDL increases as r increases. More strictly, if k is fixed where n = k + r, MTTDL increases as r increases too. For example, EC3+2 has longer MTTDL than EC3+1.
\item
If n and r are both variables, MTTDLs of two storage policies can have different relations. In Figure \ref{fig:mttdlpolicies}, EC3+2 and Replica2 have MTTDL close to each other. Their MTTDLs theoretically match at the failure rate 0.1. EC3+2 has larger MTTDL when the failure rate less than 0.1; otherwise, Replica2 outperforms EC3+2.
\end{itemize}
\section{Evaluation}
\label{sec:evaluation}

\subsection{Storage Cost}
\label{subsec:storage}

\begin{figure}[ht!]
    \centering
    \subfloat[Average redundancy count]{\includegraphics[height=1.5in, width=1.7in]{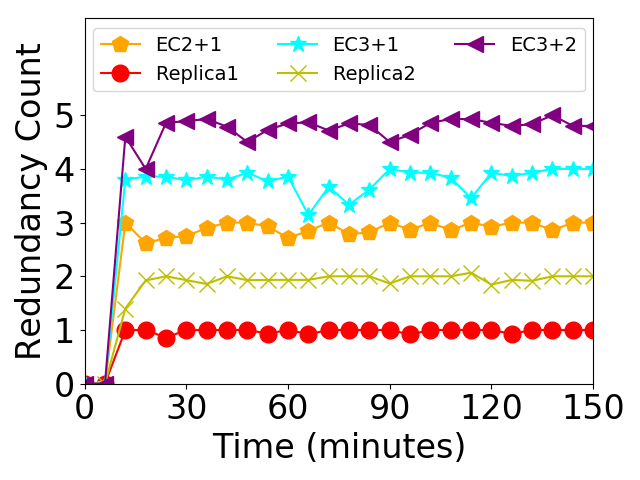}\label{fig:ecredundancycount}}
    \subfloat[Average redundancy size]{\includegraphics[height=1.5in, width=1.7in]{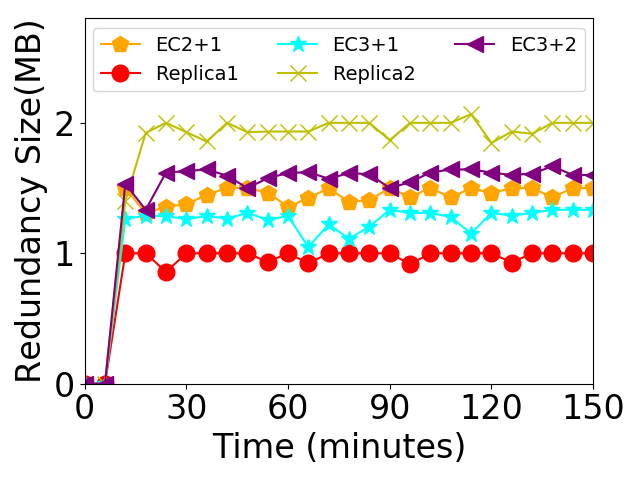}\label{fig:ecredundancysize}}
    \caption{Storage costs of erasure codes}
    \label{fig:ecstoragecost}
\end{figure}

Figure \ref{fig:ecstoragecost} shows storage costs of different policies. Figure \ref{fig:ecredundancycount} shows the average number of redundancy units and Figure \ref{fig:ecredundancysize} shows the average cache size. For example, EC3+2 stores 5 redundancy units for each cache and each unit is $\frac{1}{5}$ of the cache size. Therefore, EC3+1 stays around 1.33 MB in Figure \ref{fig:ecredundancysize}.

\subsection{Data Availability}
\label{subsec:availability}

\begin{figure}[ht!]
    \centering
    \subfloat[Cumulative temporary failures]{\includegraphics[height=1.5in, width=1.7in]{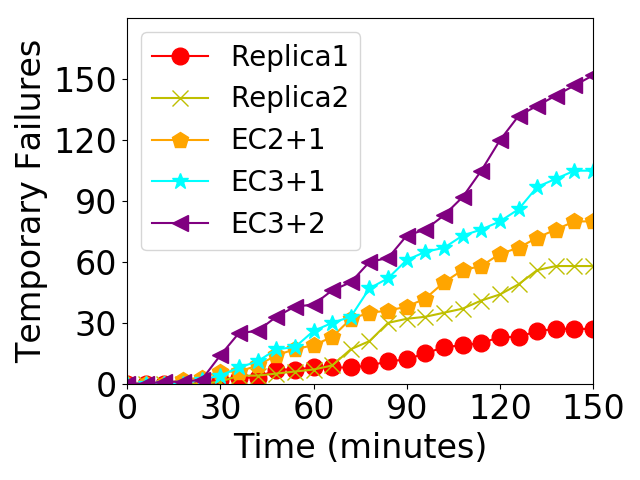}\label{fig:ectemporaryfailures}}
    \subfloat[Cumulative data loss]{\includegraphics[height=1.5in, width=1.7in]{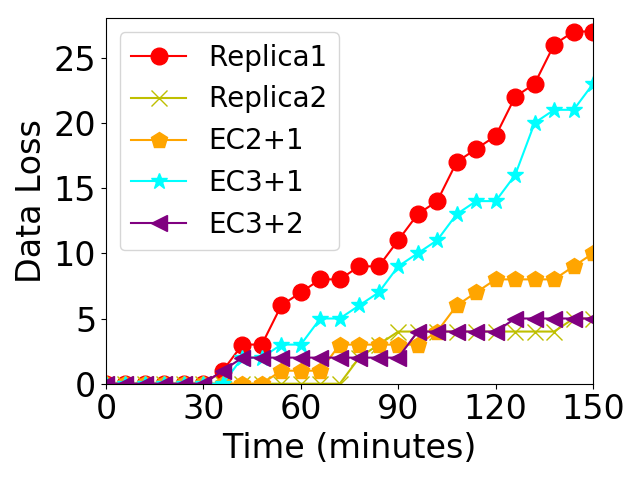}\label{fig:ecdataloss}}
    \caption{Failures of Erasure Codes}
    \label{fig:ecfailures}
\end{figure}

We use two metrics to evaluate data availability: \textit{Temporary Failures} and \textit{Data Loss}. In the test, CacheD terminations are independent of each other. The number of temporary failures is proportional to the number of CacheDs in a CacheCluster. In other words, it is proportional to the parameter n in a storage policy. Figure \ref{fig:ectemporaryfailures} proves this correlation.

It is important to know that more temporary failures do not necessarily result in more data loss. The result shown in Figure \ref{fig:ecdataloss} complies with the simulation result in Section \ref{subsec:parameters}. It is interesting to see that the data loss of EC3+2 and Replica2 are almost the same as each other in the end. A lesson learned from the figure is that both erasure codes and replication can achieve a similar data loss rate. A system can choose between them for different needs.

\subsection{Network Bandwidth}
\label{subsec:network}

\begin{figure}[ht!]
    \centering
    \subfloat[Network transfer size(MB)]{\includegraphics[height=1.5in, width=1.7in]{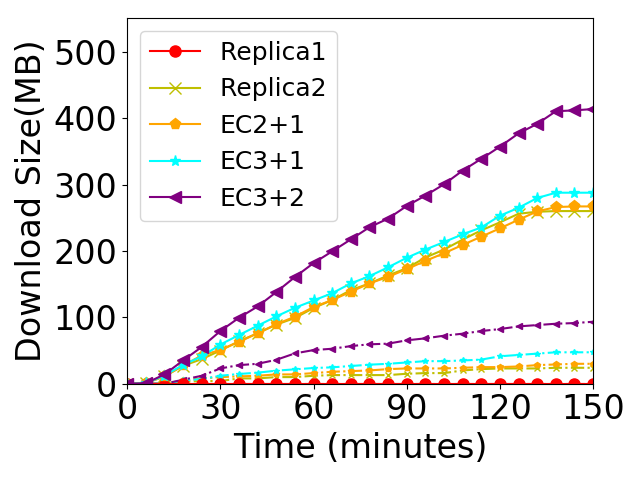}\label{fig:ecdownloadsize}}
    \subfloat[Network transfer time(s)]{\includegraphics[height=1.5in, width=1.7in]{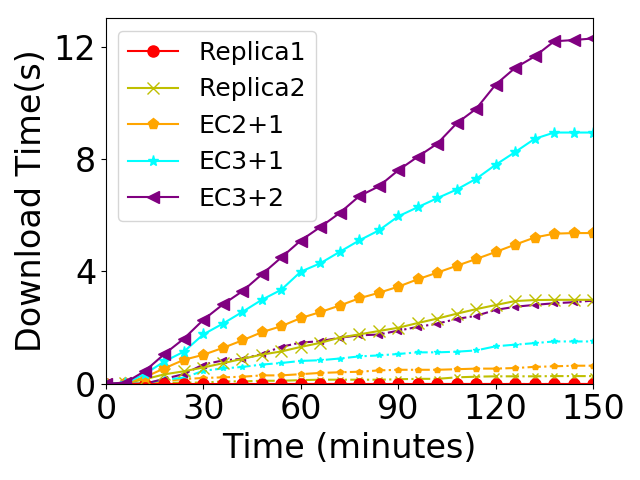}\label{fig:ecdownloadtime}}\\
    \subfloat[Network throughput(MB/s)]{\includegraphics[height=1.2in, width=1.7in]{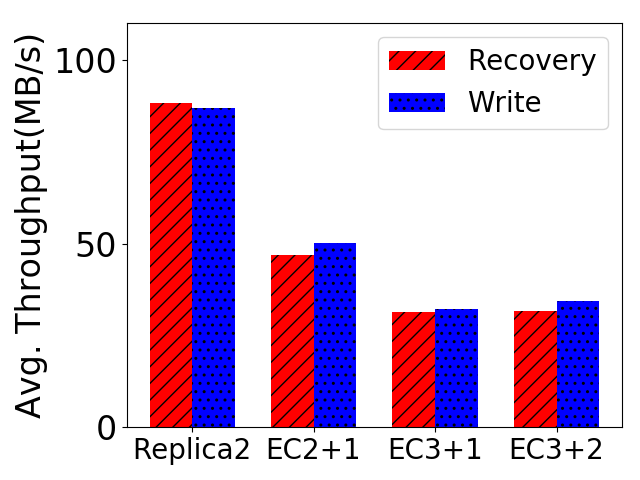}\label{fig:ecdownloadthroughput}}
    \caption{Network cost for write and recovery: the larger markers with solid lines in the figures illustrate the total network transfer time and size for different storage policies; the smaller markers with dashed lines represent the network transfers for recovery; thus, a gap between a pair of markers implies the network traffic for write operations.}
    \label{fig:ecnetworktraffic}
\end{figure}

Figure \ref{fig:ecnetworktraffic} shows the network performance on erasure codes. The dotted lines illustrate the recovery network traffic. Table \ref{tab:recoveryportion} shows the recovery portion to the overall network transfer size. One thing is worthwhile to be mentioned is that the number of redundancy units that need to be transferred on the network is always one piece less than the total number of redundancy units. This is because a CacheManager always keeps one redundancy unit on itself. In Figure \ref{fig:ecdownloadsize}, Replica2, EC2+1 and EC3+1 transfers similar amount of data over the network. EC3+2 transfers a larger amount of data than the above three policies.

\begin{table}[ht]
\centering
\caption{Recovery Network Traffic Percentage}
\label{tab:recoveryportion}
\begin{tabular}{|c|c|c|c|c|}
\hline
Storage Policy & Replica2 & EC2+1 & EC3+1 & EC3+2\\
\hline
Recovery (MB) & 24 & 30 & 47.3 & 93.3\\
Overall (MB) & 260 & 267 & 287.7 & 413.7\\
Recovery Portion & 9.2\% & 11.2\% & 16.4\% & 22.6\%\\
\hline
\end{tabular}
\end{table}

Table \ref{tab:recoveryportion} shows more redundancy units generated from a cache result in a larger portion of network traffic for recovery. This is because more temporary failures will happen in the system with a larger n. Figure \ref{fig:ecdownloadthroughput} shows the average throughput of different storage policies. As k increases, the size of each redundancy unit decreases. As a result, the size of each network transfer decreases. Although redundancy units can be transferred in parallel, the network throughput can also be affected by other sources such as TCP/IP connection setup and so on. The aggregate network throughput can be degraded due to this type of overhead. Thus, comparative results between Replica2, EC2+1 and EC3+1 prove that the larger size of each network transfer, the higher aggregate throughput a storage policy can get. Additionally, EC3+2 has a similar throughput as EC3+1 because they have the same transfer size of a redundancy unit.
\section{Proactive Redundancy Relocation}
\label{sec:proactive}

Existing research proposed proactive fault tolerance in Cloud storage systems \cite{proactiveequations}\cite{proactiveearly}. However, they mainly target drive failures in the systems. Compared with drive failures, scientific grids have a higher failure rate \cite{Zhang:2018:DJP:3219104.3229282}. In this section, we discuss a proactive approach to improve data availability in scientific grid systems.

\subsection{Design}
\label{subsec:proactivedesign}

In order to keep track of CacheWorker's age, a CacheManager keeps a hashmap in which each CacheWorker has an entry that records its booting time. The CacheManager periodically scans the map and marks any CacheWorker as PROACTIVE when it passes a pre-defined MTTDL threshold.

\begin{figure}[!ht]
\begin{center}
\includegraphics[height=1.5in, width=2.1in]{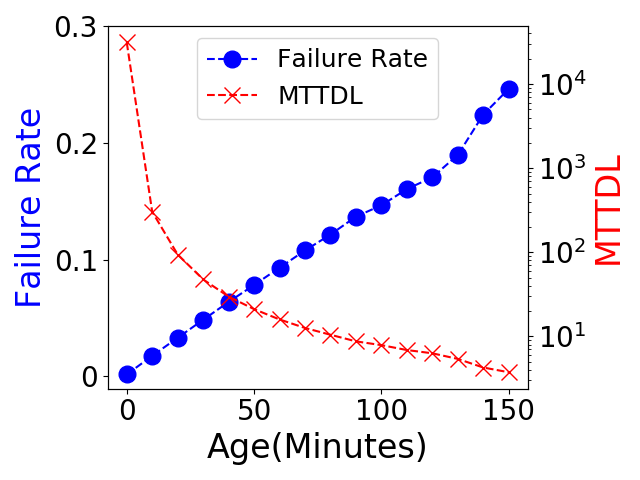}
\end{center}
\vspace*{-5mm}
\caption{A sample MTTDL for EC3+1}
\label{fig:samplemttdl}
\end{figure}

Figure \ref{fig:samplemttdl} shows MTTDL for EC3+1. We choose a threshold of 60 for MTTDL. It means once a CacheWorker's age passes 24 minutes (MTTDL is equal to 60 at the age of 24 minutes), the CacheManager marks it as PROACTIVE and starts relocating its redundancy unit to another CacheD.

\subsection{Evaluation}
\label{subsec:proactiveevaluation}

\begin{figure}[ht!]
    \centering
    \subfloat[Cumulative lifetime distribution]{\includegraphics[height=1.5in, width=1.7in]{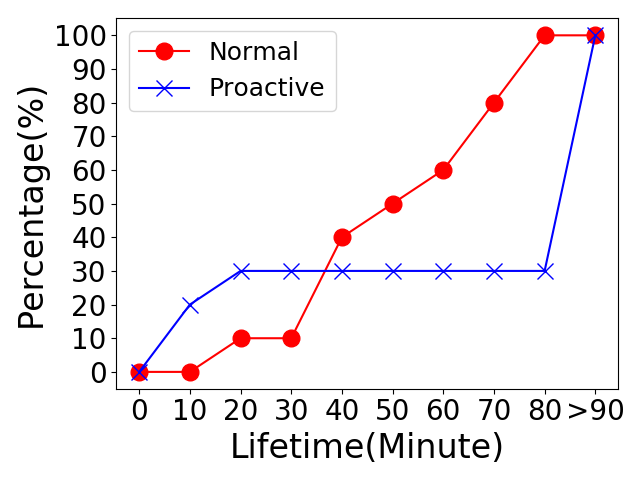}\label{fig:proactivelifetimedist}}
    \subfloat[Network traffic(MB)]{\includegraphics[height=1.5in, width=1.7in]{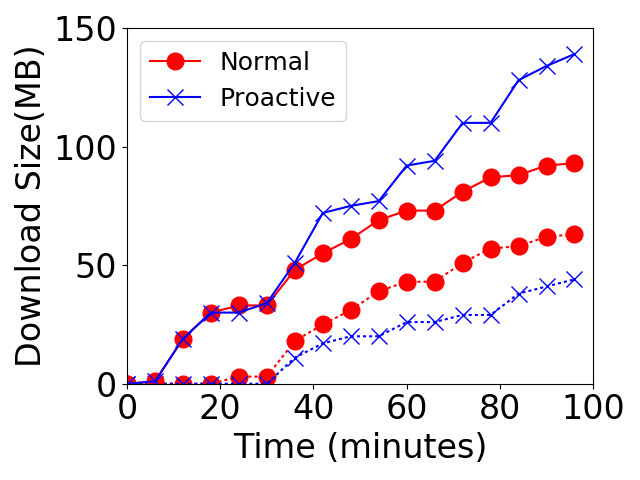}\label{fig:proactivenetworksize}}
    \caption{Comparison between proactive approach and non-proactive approach. The left figure can be explained by a few points. The red line shows the non-proactive approach: 10\% of 100 caches have the lifetime less than 20 minutes; 100\% of caches have the lifetime less than 80 minutes. The blue line shows the proactive approach: 30\% of caches have lifetime shorter than 20 minutes; 100\% of caches have the lifetime longer than the longest period we observed. Since there is a 2-minute interval for a CacheManager to check the availability of CacheWorkers, the longest period of observation is conservatively set to 90 minutes}
    \label{fig:proactive}
\end{figure}

In the previous tests, the lease period was set to 10 minutes. We extend it to 100 minutes and run a new set of experiments in which the client stores 100 caches to the cluster. Figure \ref{fig:proactivelifetimedist} shows the lifetime distribution. Without proactive relocation, none of the data survives after 100 minutes. On the other hand, the proactive approach helps to reduce the data loss to 30. Those losses happen before 24 minutes and therefore a CacheManager is not able to relocate redundancy units.

\subsection{Discussion}
\label{subsec:proactivediscussion}

The downside of proactive relocation is an increasing amount of network traffic in the system. Figure \ref{fig:proactivenetworksize} shows the network transfer size over time. All data are 1 MB and there are 100 MB scheduled to the cluster in total. The network traffic generated by recovering temporary failures is illustrated by dotted lines; solid lines represent the overall network traffic during the tests. Since proactive relocation can help to reduce temporary failures, as shown in Figure \ref{fig:proactivenetworksize} the recovery network traffic is reduced by 30\% with the proactive approach. However, the overall network traffic is increased by 49.5\%. Thus, proactive relocation offers a tradeoff of paying extra network bandwidth for higher data availability.

We only show the result of the MTTDL threshold being set to 60. The MTTDL threshold in our experiment is a configurable parameter. We do not cover the sensitivity analysis in this paper. One can expect that a smaller MTTDL threshold can offer higher data availability, yet resulting in more network traffic. Our motivation is to expose the parameter to the system in which the parameter can be adjusted due to different system performance requirements.
\section{Redundancy Localization on Network}
\label{sec:localization}

Modern scientific grid systems adopt virtualization techniques and use pilots to manage system resources and execute tasks \cite{pilotsystem}\cite{glideinWMS}. Like virtual machines in Cloud, multiple pilots can co-exist in the same physical machine. Two pilots within the same machine should have faster data transfer speed compared with those on different machines.

\subsection{Heterogeneity in Network Traffic}
\label{subsec:networkhetero}

\begin{figure}[!ht]
\begin{center}
\includegraphics[height=1.8in, width=2.5in]{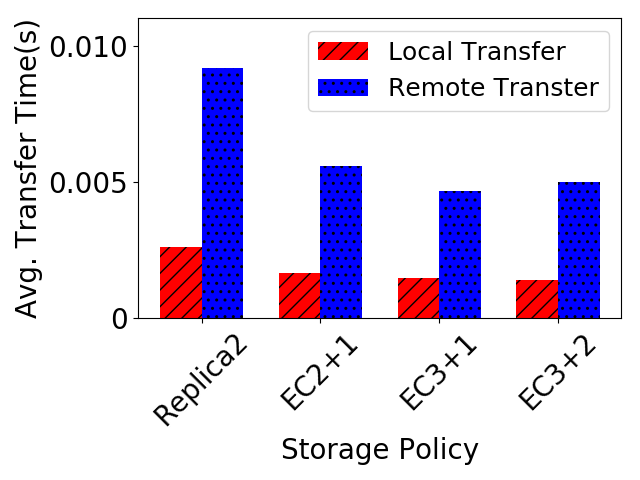}
\end{center}
\vspace*{-5mm}
\caption{Average time spent on local and remote transfer}
\label{fig:heteronetwork}
\end{figure}

Figure \ref{fig:heteronetwork} shows the average data transfer time for a redundancy unit with different storage policies. Replica2 has larger redundancy unit size (1 MB) compared with erasure codes (0.5 MB for EC2+1, 0.33 MB for EC3+1 and EC3+2). In general, local transfers only take ${\sim}30\%$ of the time of remote transfers. We only test 1 MB data stripe. Different data sizes could have different ratios but local transfers should consistently outperform remote transfers.

\subsection{Algorithm}
\label{subsec:networkalgorithm}

In order to reduce network traffic for accessing data, we introduce a parameter - \textit{LocalizationPercentage} which indicates what percentage of redundancy units in a cache should be stored within the same network domain. For example, if a cache is deployed with EC3+1 (4 redundancy units for a cache), given LocalizationPercentage equals 25\%, only 1 redundancy unit is allowed to be stored in a network domain. If LocalizationPercentage is set to 75\%, a domain can maximumly contain 3 redundancy units. If LocalizationPercentage is set to 100\%, a domain is allowed to contain all 4 redundancy units.

\begin{figure}[!ht]
\begin{center}
\includegraphics[height=2.5in, width=2in]{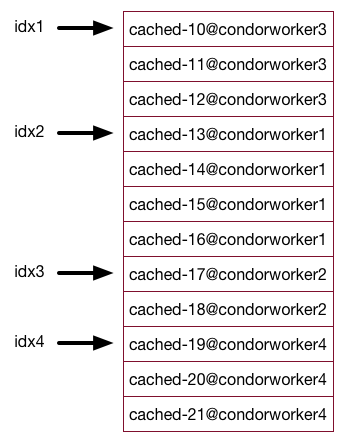}
\end{center}
\vspace*{-5mm}
\caption{Sort available CacheDs by domain}
\label{fig:idx}
\end{figure}

We divide network transfers into two cases: write path and recovery path. On a write path, all CacheDs are grouped by network domain using bucket sort. Each network domain acts as a bucket. When a CacheManager distributes redundancy units to CacheWorkers. It iterates network domains and finds the first domain that contains a sufficient number of CacheDs to store the required percentage of redundancy units. If none of the domains meet the requirement, the algorithm selects all pilots from the first domain and then move the next domain until sufficient CacheDs are selected. Figure \ref{fig:idx} shows an example. There are 12 CacheDs available on 4 VMs. They are grouped by domain. If LocalizationPercentage is set to 100\%, the EC3+1 policy selects all 4 CacheDs from condorworker1; for LocalizationPercentage of 75\%, 3 CacheDs are selected from condorworker3 and 1 CacheD is selected from condorworker1; for LocalizationPercentage of 50\%, 2 CacheDs from condorworker3 and 2 CacheDs from condorworker1 are selected; for LocalizationPercentage of 25\%, 1 CacheD is selected from each of the condorworkers.

\begin{figure}[!ht]
\begin{center}
\includegraphics[height=0.8in, width=2.5in]{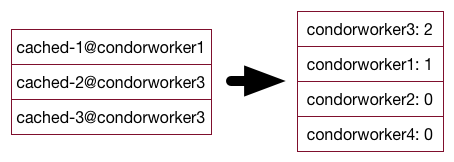}
\end{center}
\vspace*{-5mm}
\caption{Count existing CacheD domains}
\label{fig:domain}
\end{figure}

We use an example to demonstrate a recovery path. Figure \ref{fig:domain} shows an example that assumes one CacheWorker failed in a CacheCluster stored by the EC3+1 policy. In the example, there are three surviving redundancy units: one in condorworker1 and two in condorworker2. The algorithm scans surviving CacheDs and calculates their appearances by the domain name. It then sorts domain names by occurrence in descending order.

In order to reduce network traffic, the intuition is that the recovered redundancy unit should be assigned to a CacheD from a domain which contains most of the surviving CacheDs. Inspired by this idea, the algorithm sorts all surviving CacheDs by the domain rank generated from Figure \ref{fig:domain}. As a result, Figure \ref{fig:idx} shows the sorted order of all available CacheDs in the cluster according to the domain order of Figure \ref{fig:domain}. The rest of the algorithm runs in the way as a write path. In this example, if LocalizationPercentage is set to 100\% or 75\%, 1 CacheD from condorworker3 is selected. If LocalizationPercentage is set to 50\%, 1 CacheD from condorworker1 is selected. The case of LocalizationPercentage equal to 25\% should never happen according to the surviving CacheDs in Figure \ref{fig:domain}.

\subsection{Evaluation}
\label{subsec:networkeval}

\begin{figure}[ht!]
    \centering
    \subfloat[Network transfer size(MB)]{\includegraphics[height=1.5in, width=1.7in]{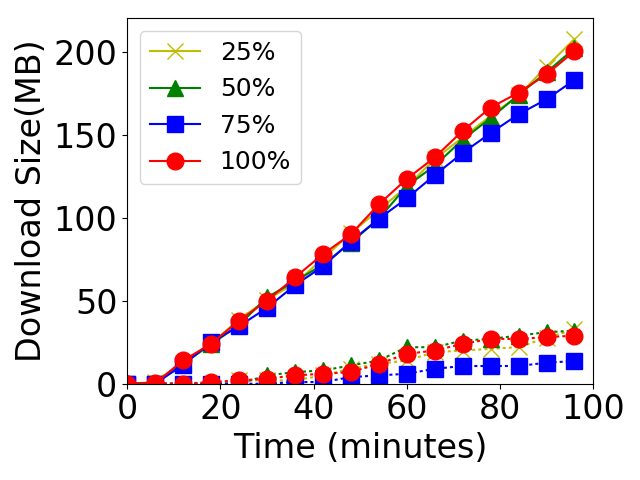}\label{fig:localdownloadsize}}
    \subfloat[Network transfer time(s)]{\includegraphics[height=1.5in, width=1.7in]{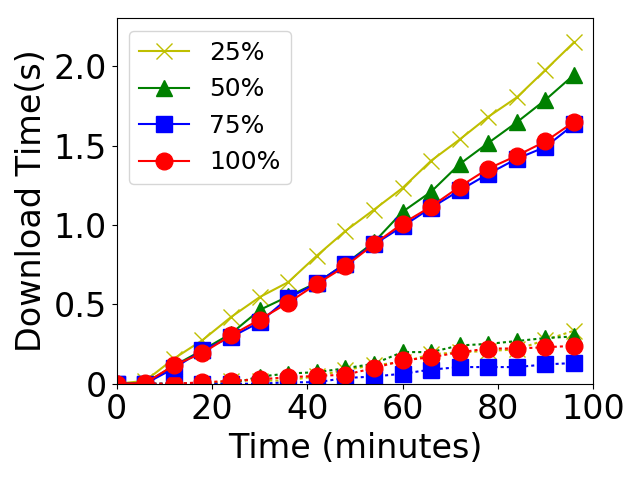}\label{fig:localdownloadtime}}
    \caption{Redundancy localization performance}
    \label{fig:localnetworktraffic}
\end{figure}

We used EC3+1 as the storage policy to evaluate LocalizationPercentage: 25\%, 50\%, 75\%, and 100\% and the results of network performance are shown in Figure \ref{fig:localnetworktraffic}. Figure \ref{fig:localdownloadsize} shows the total size of network transfer: solid lines are total network transfer and dotted lines are recovery network transfer. All tests are expected to transfer relatively the same amount of data. As shown in the figure, except for the test of 75\% that has slightly fewer recoveries and thus lower total network transfer, all tests transfer the same amount of data.

Figure \ref{fig:localdownloadtime} shows the total network transfer time during the tests. Although different localization percentages transfer the same amount of data, their times spent on network transfers are different. As the localization percentage increases, the network transferring time reduces. Since we use EC3+1 as the storage policy, each redundancy unit is only 0.33 MB. As shown in Figure \ref{fig:heteronetwork}, local transfer time for a redundancy unit is ${\sim}30\%$ of remote transfer time. As the data size increases, the performance difference between local transfer and remote transfer should increase. The system should benefit more from network localization for reducing network transfer time.

\subsection{Discussion}
\label{subsec:networkdiscussion}

The potential downside of network localization is the data transfer imbalance in the system. For example, given the localization percentage as 100\%, it is possible that at a certain period of time, all network traffic goes to the same physical machine (VM in our case).



\begin{table}[ht]
\centering
\caption{Workload variance on 4 VMs}
\label{tab:variance}
\begin{tabular}{|c|c|c|c|c|}
\hline
LocalizationPercentagey & 25\% & 50\% & 75\% & 100\%\\
\hline
VM variance & 0.094 & 0.099 & 0.101 & 0.238\\
\hline
\end{tabular}
\end{table}

Table \ref{tab:variance} shows the network imbalance on 4 VMs with different localization percentages. We grouped the total redundancy units on each VM by a 30-second interval and calculated average variance between 4 VMs over time. As the percentage increases, the redundancy variance on different VMs increases, which implies a larger traffic imbalance occurred on the network.

Besides network imbalance, node failure is another concern for the redundancy localization: if a whole data stripe is stored within the same machine, it is not resilient to machine failures. We do not address this issue in this paper. However, the algorithm can be extended to take node failure into account. The LocalizationPercentage parameter can be tuned to distribute redundancy units over wider network domains so that a cache can survive node failures.
\section{Conclusion and Future work}
\label{sec:future}

In this paper, we discussed using erasure codes to store intermediate data. An interesting observation is that there might be an erasure coding policy that can offer similar data availability as a replication policy does. Systems can choose between two policies for their own needs. We further proposed to proactively relocate redundancy to improve data availability; we also proposed redundancy localization on the network to reduce network traffic for erasure codes. Both algorithms offer configurable variables that can be tuned to benefit different system metrics. In the future, we can implement a robust storage system for intermediate data. The system can be adjusted by choosing storage policies as well as tuning a few parameters to conform to various system conditions.


\section*{Acknowledgment}

This work was supported by NSF award PHY-1148698, via subaward from the University of Wisconsin-Madison. This research was done using resources provided by the Holland Computing Center of the University of Nebraska.

\bibliographystyle{IEEEtran}
\bibliography{bib/references}

\newpage

\end{document}